\documentclass{aa}
\usepackage{graphicx}
\usepackage{epstopdf}
\usepackage{txfonts}
\usepackage{natbib}
\bibpunct{(}{)}{;}{a}{}{,}
\nonstopmode

\begin{document}

   \title{Diffuse gamma-ray emission from the Galactic center - a multiple energy injection model
}

\authorrunning{Cheng et al.}
\titlerunning{Diffuse Gamma-ray Emission}

   \author{K. S. Cheng\inst{1}\and
                 D. O. Chernyshov\inst{1,2}\and
                 V. A. Dogiel\inst{1,3}
                 }

   \offprints{K. S. Cheng, \email{hrspksc@hkucc.hku.hk}}

   \institute{Department of Physics, University of Hong Kong, Pokfulam Road, Hong Kong, China
         \and
         Moscow Institute of Physics and Technology, Institutskii lane, 141700 Moscow Region, Dolgoprudnii, Russia
         \and
             I.E.Tamm Theoretical Physics Division of P.N.Lebedev Institute, Leninskii pr, 53, 119991 Moscow, Russia
             }

  \abstract{We suggest that the energy source of the observed diffuse gamma-ray emission from the direction of the Galactic center is the Galactic black hole Sgr A*, which becomes active when a star is captured at a rate of $\sim 10^{-5} $ yr$^{-1}$. Subsequently the star is tidally disrupted and its matter is accreted into the black hole. During the active phase relativistic protons with a characteristic energy $\sim 6\times 10^{52}$ erg per capture are ejected. Over 90\% of these relativistic protons disappear due to proton-proton collisions on a timescale $\tau _{pp} \sim 10^4$ years in the small central bulge region with radius $\sim 50$ pc within Sgr A*, where the density is $\ge 10^3$ cm$^{-3}$. The gamma-ray intensity, which results from the decay of neutral pions produced by proton-proton collisions, decreases according to $e^{-t/\tau _{pp}}$, where $t$ is the time after last stellar capture. Less than 5\% of relativistic protons escaped from the central bulge region can survive and maintain their energy for $>10^7$ years due to much lower gas density outside, where the gas density can drop to $\sim 1$ cm$^{-3}$. They can diffuse to a $\sim 500$ pc region before disappearing due to proton-proton collisions. The observed diffuse GeV gamma-rays resulting from the decay of neutral pions produced via collision between these escaped protons and the gas in this region is expected to be insensitive to time in the multi-injection model with the characteristic injection rate of $10^{-5}$ yr$^{-1}$. Our model calculated GeV and 511 keV gamma-ray intensities are consistent with the observed results of EGRET and INTEGRAL, however, our calculated inflight annihilation rate cannot produce sufficient intensity to explain the COMPTEL data.
\keywords{comic rays : general - Galaxy : center - Galaxies : gamma-rays - black hole - radiation mechanisms : nonthermal
               }
            }
   
\date{Received:    }

   \maketitle
%
%________________________________________________________________

\section{Introduction}

A supermassive black hole known as Sgr $A^*$, with a mass of $\sim 3.6\times 10^6M_\odot$
\citep[for a recent review of the properties of the Galactic center (GC) black hole see][]{genkar07} is located at the GC.
In addition, there are many high energy sources harbored in this region \citep{melfal01}. For high energy radiation ranging from 511 keV annihilation lines detected by INTEGRAL \citep[e.g.][]{chusun05}, 
 in the range 1 MeV$-$30 MeV, flux was measured by COMPTEL \citep{strmos00}, 30 MeV$-$10 GeV photons by EGRET \citep[e.g.][]{mayber98}, and  TeV photons detected by Whipple \citep{kosbad04}, by CANGAROO \citep{tsueno04}, and by HESS \citep{ahaakh04} are observed from the direction of the GC.

The spatial distribution of these four energy bands is as follows: The 511 keV annihilation lines are emitted from a non-spherical symmetric extended region with about $6-8$ degrees FWHM centered at the GC. The emission appears to be diffused and does not show any clear point source in the emission region.  The annihilation flux is about $\sim 10^{43}$ ph s$^{-1}$ or $\sim 10^{37}$ erg s$^{-1}$. The MeV emission does not show any strong concentration in the direction of the GC and the emission is rather diffuse. An excess (about a factor of two above the prediction of the standard model) is found in this energy range whose origin is still unclear. The GeV source detected by EGRET is known as 3EG J1746-2851, which has an emission region around 0.5 degrees in radius with the flux $\sim 10^{37}$ erg s$^{-1}$, and the GC is at the rim of its emission region. This region is surrounded by an extended region of diffuse emission whose flux is about $\sim 10^{38}$ erg s$^{-1}$. The emission position of TeV photons with a flux $\sim 10^{35}$ erg s$^{-1}$ can be determined to be less than 10 pc and its center almost overlaps with the Galactic black hole. 

Although the emission regions of GeV and TeV photons have some overlap, it is unclear if they are related. In general, 511 keV photons are not considered to be related to both GeV and TeV photons. Furthermore, the annihilation line  is considered to be the most difficult problem to be explained. For the secondary origin of positron produced by GeV protons with nuclei of background gas this implies that the energy content in primary relativistic protons necessary to create the observed annihilation flux $\sim 10^{43}$ ph s$^{-1}$  is about $\sim 3\times 10^{54}$ erg. This amount of energy is very difficult for any known mechanisms, except gamma-ray bursts, occurring at the cosmological distance to provide. \citet{parcas05} argued that indeed the observed positrons result from the hypernova explosion, which is the progenitor of the gamma-ray bursts. However, the energy claim that gamma-ray bursts can reach
$3\times 10^{54}$erg is assumed to be isotropic emission. It is generally accepted that the emission of gamma-ray bursts is beaming and hence the energy emitted is actually two to three order of magnitudes lower than the isotropic case. 

\citet{pra06} suggests that most positrons are produced in the disk but they are transported to the bulge by the regular magnetic field. If this is true, then similar propagation should be assumed  for relativistic protons and electrons which also propagate by diffusion along magnetic field line. Their propagation in the perpendicular direction is due to random fluctuations (spaghetti-like structure). If Prantzos's model is correct we would observe extremely high fluxes of radio and gamma-ray emission from the GC. In fact the actual galactic magnetic field is quite complicated \citep{hanman06}, and as follows from radio data \citep[see][]{bec07} the derived structure of the Galactic magnetic field does not correspond to Prantzos's assumptions. \citet{masoze94} argued that the gamma-rays originating from the Galactic black hole may possibly be produced from relativistic particles accelerated by a shock in the accreting plasma. At the same time,
the gamma-rays could also come from some extended features such as radio arcs, where relativistic particles are present \citep{poh97}. \citet{marmel97} discussed in detail the gamma-ray spectrum
of GC produced by synchrotron, inverse Compton scattering, and mesonic decay resulting from the interaction of relativistic protons with hydrogen accreting onto a point-like sources (e.g. the massive black hole). However, the above models cannot produce the hard gamma-ray spectrum with a sharp turnover at a few GeV, which is observed for the GC source. Recently, \citet{okaman03} have suggested that the gamma-rays produced in the inner portion of accretion disk through the decay of neutral pions
created by p-p collisions may contribute to the gamma-rays observed by EGRET. However, their model predicted gamma-ray intensity is at least two orders of magnitude lower than the observed intensity.

\citet{cheche06} have suggested when a star is captured by a supermassive black hole at the GC, the star will be accreted into the black hole and a jet may be emitted during the accretion process. This idea comes from the fact that accreting black holes systems are seen to be accompanied with jet emission. One example of such a system is the microquasars. Studies on microquasars reveal that these objects behave very differently in their high/low state. 
In their low state, there is evidence of jet emission, although the bulk Lorentz factor of the jet is likely to be less than 2  \citep[e.g. see][]{GalFen03}. When they are in the high/soft state, there is evidence that the jet formation is greatly suppressed \citep{fencor99,GalFen03}. However, in their ``very high'' state, the jet reappears. Unlike the jet seen in the low state, the jet is very powerful and highly relativistic in the ``very high'' state of a microquasar \citep[e.g.][]{Fen03,fenmac04}.
This example shows that it is indeed possible that the transient accreting black holes are accompanied by jet emission. If jet emission also occurs during a tidal disruption event, the ejected jet will interact with the interstellar medium (ISM) and decelerate accordingly. 
\citet{wonhua07} studied the electromagnetic radiation from the jet produced from a star captured by the black hole. 
They compared the X-ray and optical data from some nearby galaxies, which are suggested to have the recent capture events. They concluded that a capture event with a characteristic jet energy $10^{52}$erg is capable of explaining the observed time dependent data. However, it is extremely difficult to observe the emission in radio because the synchrotron self-absorption has strongly suppressed the radio waves. \citet{cheche06} have assumed that the jet should consist of relativistic protons, which will gradually diffuse to a large distance away from the black hole. The proton-proton collisions can produce enough positrons to explain the observed annihilation flux of positrons from the direction of the GC.

Below we analyze the model of central black hole presented in \citet{cheche06} when positrons are ejected due to stellar capture. Since many parameters  of this process are unknown we shall try to estimate them from observed parameters of gamma-ray emission from the central region. For example in the framework of this model it gives a possibility to estimate the necessary energy release in capture processes. Our calculation in this paper is based on a multi-capture model, in which the total energy release due to multi captures in $10^7$ years is a function of the capture frequency and masses of captured stars.

The paper is arranged as follows. In Sect.~2, we describe the model, in  which relativistic protons will be ejected by the Galactic black hole when a star is captured and we introduce important timescales in the model. We estimate how many gamma-rays will be emitted from the central compact region and from the larger low density region. In Sect.~3, we outline the model calculations. In Sect.~4, we present our numerical results and compare our model results with the observed high energy radiation data from GC. In Sect.~5, we present a brief discussion.

\section{Model description}

The rate at which a massive black hole in a dense star cluster tidally disrupts and swallows stars has been studied extensively \citep[e.g.][]{hil75,bahwol76,ligsha77}. Basically when a star trajectory happens to be sufficiently close to a massive black hole, the star would be captured and eventually disrupted by tidal forces. After a dynamical time-scale (orbital time-scale), the debris of a tidally disrupted star will form a transient accretion disk around the massive black hole, with a radius typically comparable to the tidal capture radius \citep{ree88}. Rees also argued that most of the debris material is
swallowed by a black hole with a mass $\sim 10^6 M_{\odot}$ on a timescale of $\sim 1$ yr for a thick hot ring, or $\sim 10^2$ years for a thin cool disk. The more quantitative description will be given later. The capture rate is essentially a problem of loss-cone diffusion-diffusion in angular momentum rather than energy.
By assuming a Salpeter mass function for the stars, \citet{syeulm99} estimated the capture rate in our Galaxy as $\sim 4.8 \times 10^{-5}$ yr$^{-1}$ for main sequence stars and $\sim 8.5 \times 10^{-6}$ yr$^{-1}$ for red giant stars, respectively. On the other hand, \citet{magtre99} used the dynamical models of real galaxies by taking into account the refilling of the loss cone of stars on disruptable orbits by two-body relaxation and tidal forces in non-spherical galaxies, but they obtained a higher capture rate $\sim 10^{-4}$ yr$^{-1}$. Therefore, the actual capture rate is sensitively dependent on the assumed mass function of stars, the stellar evolution model used, the radius and mass of the captured star, the black hole mass and the internal dispersion velocity of stars ($\rm {v_s}$) around the black hole. For example, based on the theory of \citet{cohkul78}, \citet{chelu01} have shown that the capture rate of stars by the massive black hole is proportional to $M_{bh}^{2.33}n_*^{1.6}\rm {v_s}^{-5.76}$, where $M_{bh}$, and $n_*$ are the mass of the black hole and the star density in the star cluster around the black hole respectively. They obtained a longer capture time $\sim 10^6$ years, by taking $\rm {v_s} =10^2 \rm
{km/s}$ and $M_{bh}=3.6 \times 10^6 M_{\odot}$. Therefore the capture time for a main sequence star with mass $\sim 1M_{\odot}$ could range from several tens of thousands of years to several hundreds of thousands of years.
 It is very important to note that the correct prediction of capture rate is a very difficult task. Based on the observations of nearby galaxies, \citet{fer02}, \citet{gebric02} and \citet{tregeb02} have given some 
simple relations between the black hole mass and the velocity dispersion as $M_{bh} \sim \rm{v_s}^{4.58 \pm 0.52}$ and $M_{bh} \sim \rm{v_s}^{4.02 \pm 0.32}$ respectively. 
If we substitute these simple relations into the formula derived by Cheng \& Lu, it produces a totally unreasonable capture rate. However, \citet{barho04} have shown that such simple formulae would break down in the mass scale like the black hole in GC.
Although the stellar capture rate is difficult to be determined theoretically, there are five X-ray flare events observed in nearby normal galaxies, which are believed to be the consequences of the stellar
capture \citep{donbra02,halgez04,kom06}.
Based on these observed events the average capture rate per galaxy is about $\nu_{cap}\sim 10^{-5}$ yr$^{-1}$ (below $\tau_{cap}=1/\nu_{cap}$). 

When a star comes within the capture radius, which is given by
\begin{equation}
R_T\approx 1.4\times 10^{13} M_6^{1/3}m_*^{-1/3}r_*\,\,\rm{ cm},
\end{equation}
where $m_*=M_*/M_\odot$, $M_6=M_{bh}/10^6M_\odot$,
$r_*=R_*/R_\odot$, the star will be captured by the black hole \citep{ree88,phi89}. 
According to the theoretical predictions, the flare results from the rapid release of gravitational energy as the matter from the disrupted star plummets toward the black hole. For t$> t_{peak}$, the accretion rate
evolves as, \citep{ree88,phi89},
\begin{eqnarray}
\dot{M}\sim
\frac{1}{3}\frac{M_*}{t_{min}}\left(\frac{t}{t_{min}}\right)^{-5/3},
\label{rees}
\end{eqnarray}
where  ${M_*}$ and $R_*$  are the mass and the radius of the captured star, respectively and $t_{peak}\sim 1.59 t_{min}$, $t_{min} \approx 0.2 \left(\frac{M_{\odot}}{M_*}\right)\left(\frac{R_*}{R_{\odot}}\right)^{3/2}\left(\frac{M_{bh}} {10^6M_{\odot}}\right)^{1/2}\,\,\mbox{yr}$ is the characteristic time for the debris to return to the pericenter \citep{luche06}.
 Recently \citet{yuamar02} studied jets emission from Sgr A*. They suggested that the Chrandra observed features of X-rays from the vicinity of Sgr A* can be explained in terms of a coupled jet plus accretion disk model. The observed radiation is mainly emitted by the electrons in ADAF disk and in the jet. However, it is not known if the energy of the jet will be carried away by protons or electrons. Since the inertia of protons is much larger than that of electrons, it is logical to assume that the energy of the jet is mainly carried by protons. The energy distribution of the protons in the jet is usually assumed to be a power law and the index is taken to be 2-3 \citep{yua07}. \citet{yuacui05} studied how much mass is carried away by the jet in  the black hole system, and concluded that it is typically 1\%-10\%. 

\citet{cheche06} estimated that the maximum accretion energy carried away by relativistic protons is given by
\begin{eqnarray}
\Delta{E_p} \sim 6 \times 10^{52}(\eta_p/10^{-1}) (M_*/M_{\odot}) \mbox{\,\,erg},
\end{eqnarray}
where $\eta_p$ is the conversion efficiency  from accretion power ($\dot{M}c^2$) into the the energy of jet motion.

A number of timescales are important for our model; the proton-proton collision time scale
\begin{equation}
\tau_{pp} = (n\sigma_{pp}c)^{-1}\sim 3\times 10^7 n^{-1} \rm{\,\,  yr},
\end{equation}
the diffusion time scale
\begin{equation}
\tau_{d} = d^2/6D \sim 10^7 (d/500 \rm{\  pc})^2(D/10^{27}\rm{\  cm}^2\rm{ s}^{-1})\rm{\,\,  yr},
\end{equation}
and the ionization cooling time scale for the relativistic charged particle
\begin{equation}
\tau_{cool} = 10^8 (E/\rm{ GeV})n^{-1} \rm{\,\,  yr}.
\end{equation}
Here $n$ is the gas density, $\sigma_{pp}$ is the p-p collision cross-section, $d$ is the distance from the source, $D$ is the diffusion coefficient and $E$ is the energy of charged particle.

It is very difficult to explain the extended spatial distribution of the annihilation emission around the GC unless the sources of positrons are more or less uniformly distributed in the bulge as there is no such problem of propagation \citep[e.g.][]{wanpun06,weishr06}. The problem is that we do not know how positrons with energies below 100 MeV propagate through the interstellar space.  From observations, 
we can derive {\it average values} of the diffusion coefficient which, in principle, differ from each other depending on the analyzed spatial region. One can find these estimations in \citet{berbul90,strmos98,strmos00}. Thus, from radio and gamma-ray emission from the Galactic halo whose semithickness is about several kpc, one can  show that cosmic rays with energies above 100 MeV propagate by diffusion with the coefficient in scales of the Galactic halo of about $D_h\sim 3\cdot 10^{28}-10^{29}$ cm$^2$s$^{-1}$. From the cosmic ray chemical composition which is determined 
by particle propagation inside the gaseous disk one can find the value about $D_g \sim 3\cdot10^{27}- 3\cdot 10^{28}$ cm$^2$s$^{-1}$, i.e. smaller than in the halo. 
Similar values for the diffusion in the local Galactic medium were derived from the anisotropy of high energy cosmic rays emitted by nearby supernova shells. Besides, these estimations are strongly dependent on whether the values of D are spatially or energy dependent as well as if there are other mechanisms of cosmic ray transport in the Galaxy.

It is unclear, of course, whether the average characteristics of cosmic ray propagation in the disk or in the halo can be extrapolated onto the region of the Galactic bulge in order to describe distribution of positrons with much smaller energies.
However, rather simple estimates  for the spatial diffusion coefficient of MeV positrons give only the value, $D \sim 10^{27}$ cm$^2$s$^{-1}$ \citep[e.g.][]{jeakno06}. In the framework of our model we can argue that positrons ejected from a central region of the GC should propagate over the distance about several hundred pc during the time of their thermalization ($\sim$10 million years  in order to satisfy the observations).

The gas density distribution in the GC is complicated. According to \citet{jeakno06}, the bulge region inside the radius $\sim230$ pc and height 45 pc contains $7\times 10^7M_{\odot}$. A total of 90\% of this mass is trapped in small high density clouds (as high as $10^3$ cm$^{-3}$) while the remaining 10\% is homogeneously distributed with the average density $\sim 10$ cm$^{-3}$. In the 500 pc region the average density will drop to $\sim (1-3)$ cm$^{-3}$.

When relativistic protons are ejected after the stellar capture, they take $\sim 10^5$ years to leave the central high density region ($\sim 50$ pc)  if the diffusion coefficient of protons near the GC is about $10^{27}$ cm$^2$s$^{-1}$ i.e. as in the Galactic Disk \citep{berbul90}.
The proton collision timescale is only $3\times 10^4$ years. Therefore, most proton energy will be converted into pions, which quickly decay to photons, electrons and positrons.

The most plausible observed gamma-ray intensity must be of the order of
\begin{eqnarray}
L_{\gamma}(50 \rm{\  pc})&\approx& \frac{\eta_{\pi}\Delta E_p}{\tau_{cap}}
e^{-\tau_{cap}/\tau_{pp}} \nonumber\\
&\approx &6\times 10^{37}
(\eta_{\pi}/10^{-1}) (\eta_p/10^{-1}) (M_*/M_{\odot}) \rm{\,\, erg/s},
\label{g50}
\end{eqnarray}
where $\eta_{\pi}$ is the conversion efficiency from protons to neutral pions.

The clumpy structure of the gas in the central 50 pc region is essential for estimates of the lifetime of primary protons because the primary protons may disappear after the first penetration into dense clouds if the gas density there is high enough. However, since the fraction of protons leaving the high density region is $e^{-(\tau_{d}/\tau_{pp})}\sim e^{-3}\sim 0.05$ then the number of clouds should be large enough and the approximation of average density is completely acceptable.

These 5\% of primary protons leaving the central region can propagate  by diffusion through the 1 cm$^{-3}$ interstellar gas to the distance about 500 pc. Since the p-p collision time in 500 pc is $10^7$ years, the proton injection rate into this region is almost constant in comparing with the diffusion timescale. Therefore the gamma-ray emission intensity from 500 pc region is almost constant and the gamma-ray power is given by
\begin{eqnarray}
L_{\gamma}(500\rm{\  pc})&\approx &0.05\frac{\eta_{\pi}\Delta
E_p}{\tau_{cap}} \nonumber\\
&\approx &3\times 10^{38} (\eta_{\pi}/10^{-1})
(\eta_p/10^{-1}) (M_*/M_{\odot}) \rm{\,\,  erg/s}. \label{g500}
\end{eqnarray}

It is very important to note that most of positrons are created in this high density region. However, they may propagate mainly through the intercloud medium because of the screening effect due to  MHD waves excited near dense molecular clouds \citep{skistr76,dogsha85,padsca05}. 
It is unclear at which precise energies of positrons this effect is essential but if these waves are excited by subrelativistic nuclei then positrons with energies as high as (and below) 300$-$600 MeV cannot penetrate into the clouds \citep{mor82}.
 This means that a significant part of secondary positrons even  with relativistic energies is cooled down in the intercloud medium only. 
The cooling time scale of positrons  is as large as $\sim 10^7$ years.

The expected positron annihilation rate in this case is given by
\begin{equation}
\dot{N_e^+}\approx \frac{\Delta E_p}{m_pc^2\tau_{cap}} \approx
2\times 10^{43}(\eta_p/10^{-1}) (M_*/M_{\odot}) \rm{ \ s}^{-1},
\label{an500}
\end{equation}

The estimates (\ref{g50}) and  (\ref{g500}) are completely
consistent with the EGRET data \citep{mayber98} and the estimate
(\ref{an500}) with the observed results of INTEGRAL \citep[e.g.][]{chusun05}.

\section{Model calculations}
The detailed description of model calculations was given in \citet{cheche06}. Here we summarize the
calculation procedure as follows.

Since most accretion energy will be released over a very short timescale, we take the source function of protons as
\begin{equation}
Q(r,E_p,t)=A(E_p)\delta({\bf r})\delta(t),
\end{equation}
where $
A(E_p)\propto{{E_p+M_pc^2}\over{(E_p^2+2M_pc^2E_p)^{(\gamma_0+1)/2}}}$
for the power-law momentum injection spectrum and the
spectral index $\gamma_0$ is taken to be between 2-3 \citep[cf.][and reference hereafter]{berbul90}. 

Since the acceleration processes in the jet are still unclear, it is not easy to determine the maximum energy of protons in the jet. According to the TeV observations from AGNs \citep[e.g. Mkn 421 and Mkn 501, cf.][]{wee04}, if we assume that these TeV photons result from proton-proton collisions, the energy of protons in the jet must be at least over 10 TeV. In fact, TeV photons have also been observed from the vicinity of Sgr A* \citep{kosbad04,tsueno04,ahaakh04}, it has been suggested that these TeV photons result from the p-p collisions and the relativistic protons are ejected from the Galactic black hole \citep[e.g.][]{ahaner05a,ahaner05b,luche06}. However, the exact values of the maximum protons are not important in our problem. In order to fit the EGRET data, the spectral index of the proton spectrum is required to be close to 3. Therefore most proton energy is at $\sim$GeV.

The spatial distribution of the protons can easily be derived from the well-known equation of cosmic ray propagation,
\begin{equation}\label{eqp}
{{\partial n_p}\over{\partial t}}-\nabla (D \nabla n_p) +
{{\partial }\over{\partial
t}}\left({{dE}\over{dt}}n_p\right)+{n_p\over {\tau_{pp}}} = Q({\bf r},E_p,t).
\end{equation}
Here $D$ is the diffusion coefficient, 
\begin{equation}
{{dE}\over{dt}}=-{{2\pi e^4n}\over{mc\beta(E)}}\ln\left({{m^2c^2W_{max}}
\over{4\pi e^2\hbar^{2} n}}\right)
\end{equation}
 is the rate of ionization losses and $\tau_{pp}$ is the characteristic time of p-p collisions.

Neutral pions are produced in p-p collisions and they will decay almost immediately to high energy photons. The emissivity of the photons produced by $\pi^{0}$-decay ($q_{\gamma}$) can be
calculated from corresponding references for this and other equations used for calculations of gamma-ray and secondary particles which one can find in \citet{cheche06}:

\begin{equation}\label{eq_gamma_pi_0}
q_\gamma(E_\gamma, t) = 2\int\limits_{E_{\pi min}}^{\infty}
\frac{q_\pi(E_\pi,t)}{\sqrt{E_\pi^2 - m_\pi^2c^4}}dE_\pi
\end{equation}
where $q_{\pi} = \int_{E_{p}}N(E_{p})v_pn_{H}d\sigma(E,E_{p})$,
$d\sigma(E,E_{p})$ is the differential cross-section for pions,
$E_\gamma$ and $E_\pi$  are the energy of the emitted photon and the 
decaying pion respectively, $E_{\pi min}=E_\gamma+\frac{m_\pi^2c^4}{E_\gamma}$

Inelastic $p-p$ collisions produce two charged pions for every neutral pion. 
These charged pions quickly decay into muons, which in turn decay into positrons
and electrons, with a resulting emissivity
\begin{eqnarray}\label{ne}
q_e(E_e)&=& n_{H}\frac{m_{\pi}^{2}c}{m_{\pi}^{2}-m_{\mu}^{2}}\nonumber\\
&&\int\limits _{E_\mu^{min}}^{E_\mu^{max}} dE_\mu \frac{dP}{dE_e} \int\limits
_{E_\pi^{min}}^{E_\pi^{max}} \frac{dE_\pi}{\beta_\pi E_\pi}
\int_{E_{th}(E_\pi)} dE_p n_p \frac{d\sigma(E_\pi,E_p)}{dE_\pi}
\end{eqnarray}

We use the generalized Fokker-Planck equation to calculate the positron distribution function $f$  written in dimensionless variables $p=p/\sqrt{mkT}$ and $t=\nu_0t$ is
\begin{eqnarray}
\frac{\partial f}{\partial t}+{{vn(\sigma_{an}+\sigma_{ce})}\over
\nu_0} f&-&q_e(p,{\bf r})\nonumber\\
& =& \frac{1}{p^{2}}\frac{\partial} {\partial
p}\left[{ A}(p)\frac{\partial f(p)}{\partial p}+{B}(p) f\right]\,,\label{eq_pos}
\end{eqnarray}
where $(\sigma_{an})$ and $(\sigma_{ce})$ are the cross-sections for inflight annihilation and charge-exchange processes respectively, $q_e$ describes the distribution of sources emitting fast positrons, $\nu_0={{2\pi
\bar{n}_{e}c^{2}r_{e}^{2}m_{e}}\over{\sqrt{m_{e}kT_{x}}}}$,
${A}(p)=p^{2}\left[-\left(\frac{dp}{dt}\right)_{ion}\frac{\gamma}{\sqrt{\gamma^{2}-1}}
\sqrt {\frac{kT_{x}}{m_{e}c^{2}}}\right]$ and ${B}(p)=
p^{2}\left[-\left(\frac{dp}{dt}\right)_{ion}-
\left(\frac{dp}{dt}\right)_{synIC}-\left(\frac{dp}{dt}\right)_{brem}\right]$.

The high energy gamma-rays ($>30$ MeV) are mainly produced by the decay of neutral pions, whereas the lower energy gamma-rays ($<30$ MeV) are produced by in-flight annihilations of positrons. The
in-flight differential spectrum of the $\gamma$-rays produced by annihilation of a positron on the ambient electrons with density
$n_{e}$:
\begin{eqnarray}
q_{an}(\varepsilon) = \frac{\pi r_{e}^{2}cn_{e}}{\gamma_{+}p_{+}}\left[
\left(\frac{\varepsilon}{\gamma_{+}+1-\varepsilon} \right. +
\frac{\gamma_{+}+1-\varepsilon}{\varepsilon}\right) + \\
2\left( \frac{1}{\varepsilon}+\frac{1}{\gamma_{+}+1-\varepsilon}\right) -
\left.\left(\frac{1}{\varepsilon}+\frac{1}{\gamma_{+}+1-\varepsilon}\right)^{2} \right] %%@
\nonumber
\end{eqnarray}
where $\gamma_{+} = E_{+}/m_{e}c^{2}$ is the Lorentz-factor of the positron, $p_{+} = %%@
\sqrt{\gamma_{+}^{2}-1}$ is the dimensionless momentum of
positron and $\varepsilon = E/m_{e}c^{2}$ is the dimensionless photon energy. The energy %%@
spectrum of in-flight annihilation
\begin{equation}
N_{an}(\varepsilon) = \int\limits_\gamma^\infty
q_{an}(\gamma_{0},\varepsilon)n_{e}(\varepsilon)d\gamma_{0}
\end{equation}
The lower integration limit can be obtained from the kinetic equations and
equals
\begin{equation}
\gamma(\varepsilon) = \frac{\varepsilon^{2} + (\varepsilon-1)^{2}}{2\varepsilon-1}.
\end{equation}

\section{Numerical calculations}

\subsection{Energy input constraints of multi-injection model}

From Eq.~(5), we see that the energy carried away by the relativistic protons can vary from capture to capture depending on the mass of the captured star and the efficiency of converting accretion energy into outflow of relativistic protons. It ranges from $10^{52}$ erg for $\eta_p \sim 0.01$ and $M_*/M_{\odot}\sim 1$ to $10^{54}$ erg for $\eta_p \sim 0.1$ and $M_*/M_{\odot}\sim 10$. In numerical calculations, we can only assume a mean energy injection in Eq.~(10). To constrain the injection energy, we used the observed electron/positron annihilation intensity to constrain the injection energy as follows:
 In $p-p$ collisions only 4\% of initial energy of proton is transferred to secondary electron which gives the energy of positrons of about 40 MeV for a 1 GeV primary proton, and the possibility of the reaction $p + p \rightarrow \pi^+ + ...$ is about 1/3. 
 Consequently, to maintain the annihilation flux at the observed level we should supply about $10^{43}\times 40 \times \frac{1}{0.04} \times 3 = 2.25\times 10^{46}$ MeV/s or $3.6\times 10^{40}$ erg/s. Since the time of positron thermalization  is about $3 \times 10^{14}$ s we need the total energy in relativistic protons $\sim 10^{55}$ erg, in order to produce the observed annihilation flux in the event of a single
eruption. This amount of injection energy is possible by capturing a massive star with $\sim 30M_{\odot}$ plus high conversion efficiency. However, the most natural explanation is a multiple capture scenario (multiple energy injection model). 
Since the capture rate is $\sim 10^{-5}$ yr$^{-1}$, there would be about 100 captures in the past $10^7$ years and each of them only requires $\sim10^{53}$ erg, which is the most typical value in Eq.~(5). This non-stationary model also gives a natural explanation as to why the gamma-ray intensity emitted from the central high density region
is so low.

%                                     Two column figure (place early!)
%______________________________________________ Gamma_1 (lg rho, lg e)
   \begin{figure*}
   \centering
\includegraphics[width=18cm]{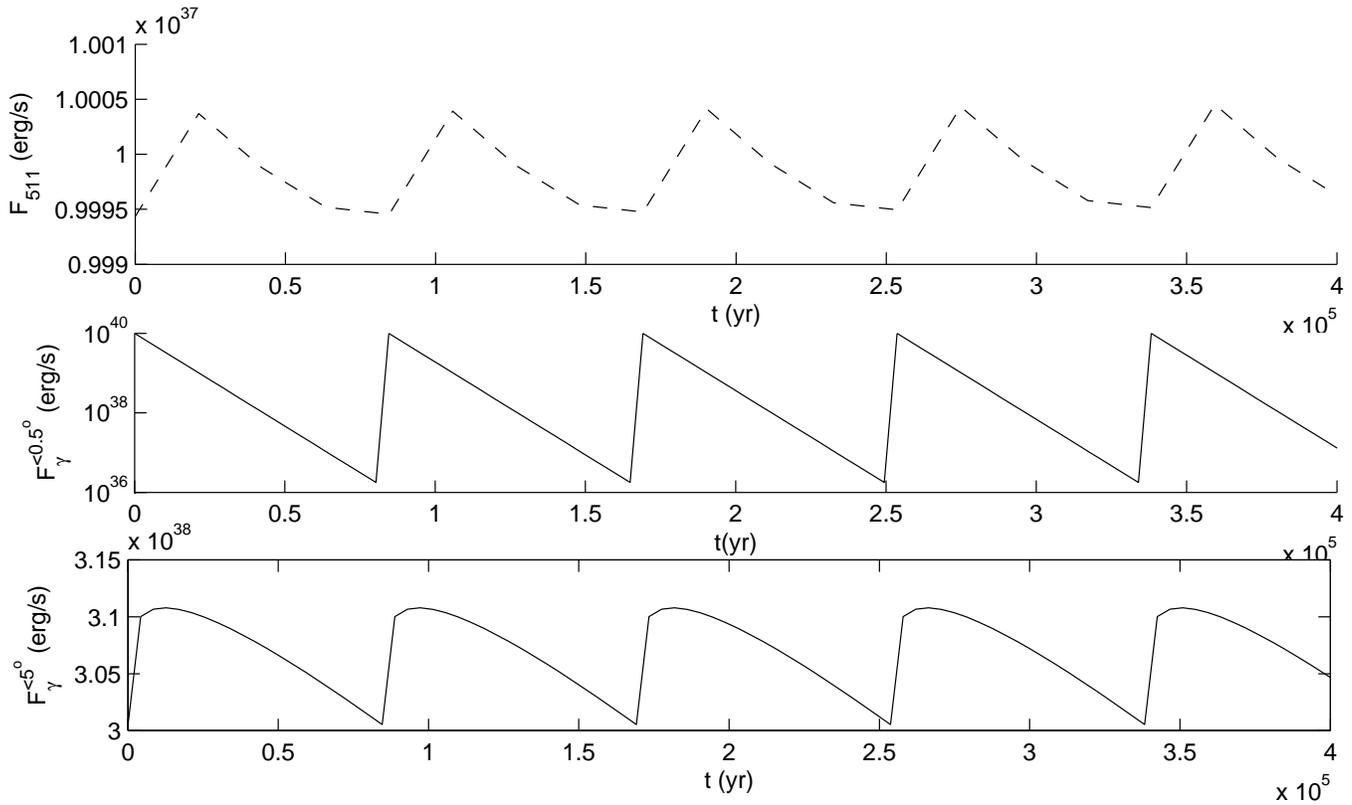}
   \caption{The time variations of
annihilation emission, gamma-ray flux from the central high
density region (50 pc), and the flux of gamma-rays produced by protons
 escaping from the central core into the low density region (500 pc).}
              \label{fluxes}%
    \end{figure*}

In Fig.~1, we can see that both annihilation flux and gamma-ray flux in large regions are almost constant whereas the gamma-ray flux in the central high density region is much more sensitive to time. Furthermore the energy requirement of the multi-injection model is much less than that of the single capture model due to accumulation of positrons in the thermal energy region (Fig.~2). Our calculations show that this release should be about $5\times 10^{52}$ erg i.e. a capture of one solar mass stars once
a period $\tau_t\sim 10^5$ years which is quite enough to generate the observed annihilation emission. This value is much smaller than the energy release in primary protons required for a single capture model (Cheng et al.~2006) which is about $10^{55}$ erg for the intercloud gas density $\sim 1$ cm$^{-3}$.

%                                     Two column figure (place early!)
%______________________________________________ Gamma_1 (lg rho, lg e)
   \begin{figure*}
   \centering
\includegraphics{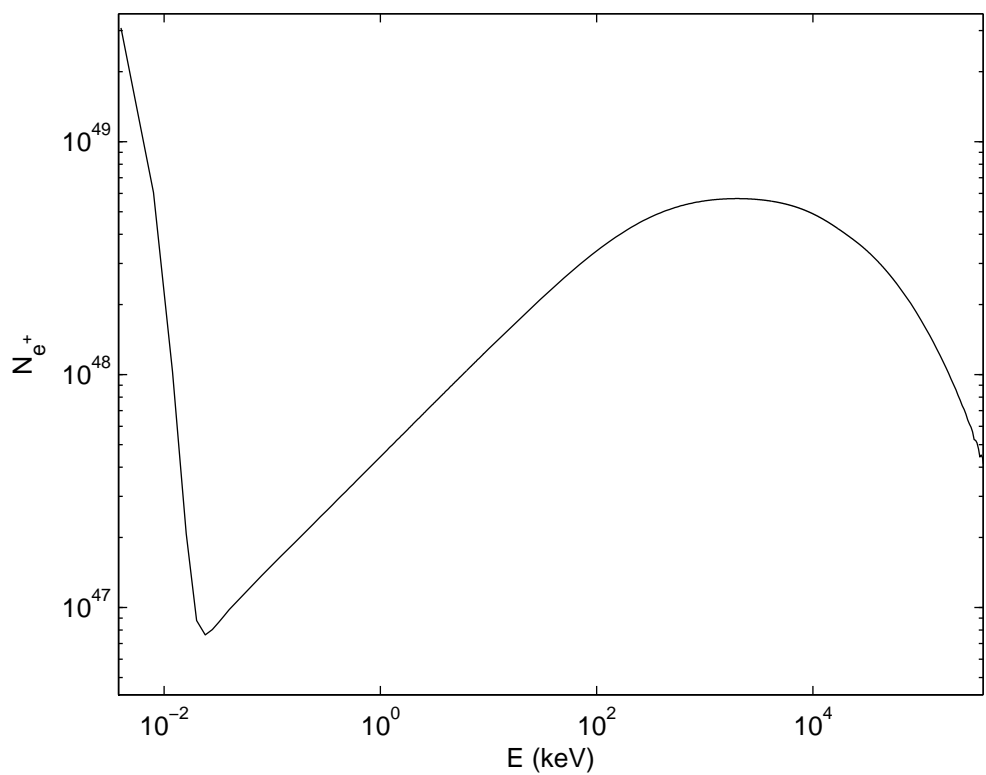}
   \caption{The steady state distribution of positrons in the low density region %%@
(500 pc).}
              \label{positron}%
    \end{figure*}

\subsection{MeV-GeV gamma-rays}
In Fig.~3, we calculated the wide band spectrum. In this paper we compare the model calculations with the observed data of EGRET \citep{mayber98} and COMPTEL \citep{strdie05}. In \citet{strdie05} they summarized the data of EGRET from 30 MeV$-$30 GeV, COMPTEL from $1-10$ MeV and INTEGRAL from 100 KeV$-$0.5 MeV for the more extended region. The input energy in protons is chosen to reproduce the observed annihilation line flux. We can see that the model curve is consistent with the EGRET data in $5^{\circ}$ radius but is substantially lower than the COMPTEL, which means that the
in-flight annihilation of our multi-capture model cannot produce the observed $1-10$ MeV photons whereas the neutral pions decay is capable of explaining the GeV photon emission. Based on results of \citet{beayuk06}, \citet{tot06} suggested that the model proposed by \citet{cheche06} will produce too
much $1-10$ MeV gamma-rays through in-flight annihilation and hence does not satisfy the observed data of COMPTEL. However, with the detail calculations by solving the proper kinetic equation, we can
show that the in-flight annihilation will not over produce $1-10$ MeV. On the contrary, calculations in the framework of \citet{cheche06} cannot produce enough $1-10$ MeV gamma-rays to explain the COMPTEL data as shown in Fig.~3. We want to emphasize that most positrons are produced in the central high density region ($\sim 50$ pc). Then they take $\sim 10^7$ years in the low density region
($\sim 500$ pc) to become thermalized positrons, which can capture electrons to form positronium. Our model injection of positrons in the low density region is equivalent to a constant injection of
positrons with energy lower than 10 MeV, which does not violate the constraint concluded by \citet{beayuk06}.

However, in the energy range above 10 MeV, the in-flight annihilation emission is quite significant and gives the main contribution at $\sim 30$ MeV. It is interesting to note that at this energy the in-flight flux exceeds the contribution from inverse Compton scattering of relativistic electrons which is essential in the Galaxy especially at relatively high latitudes \citep{doggin89,strmos00}.

%                                     Two column figure (place early!)
%______________________________________________ Gamma_1 (lg rho, lg e)
   \begin{figure*}
   \centering
\includegraphics{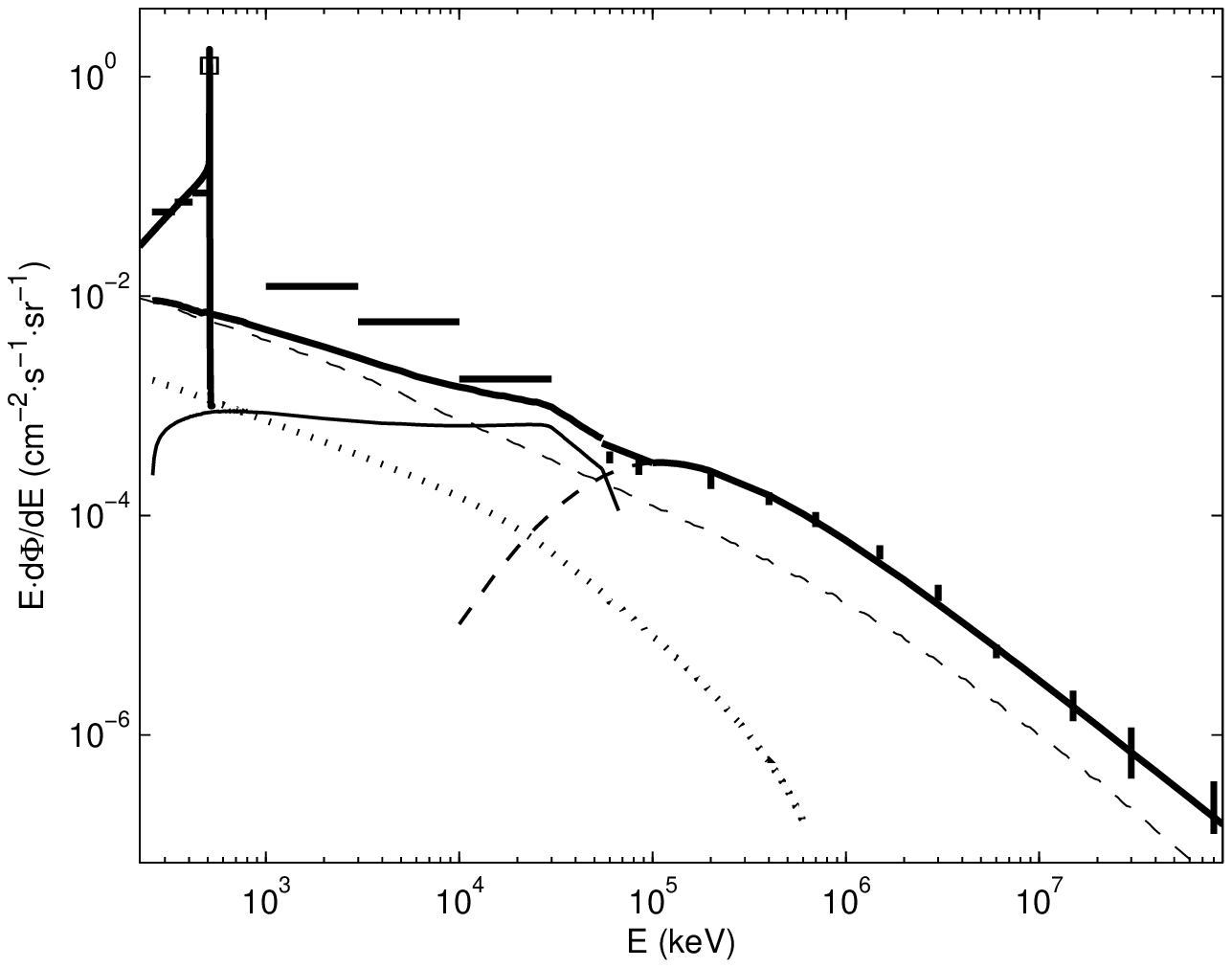}
   \caption{The theoretical broad
gamma-ray band spectrum and the observed data of EGRET, COMPTEL
and INTEGRAL. Gamma-rays with energies from 30 MeV-100 GeV (EGRET
range) are dominated by the decay of neutral pions (dashed thick
line), gamma-rays with energies from 1-10 MeV (COMPTEL range) are
dominated by in-flight annihilation (thin solid line) and inverse
Compton scattering of relativistic electrons (thin dashed line);
the IC data ware taken from \citet{strdie05}. The gamma-rays
with energies about 500 keV are dominated by the electron-positron
annihilation via positronium, in which two photon-decay produces
the line spectrum and the three-photon decay produces the
continuum. The dotted line presents the bremsstrahlung radiation
of secondary electrons.}
              \label{gammaray}%
    \end{figure*}

\section{Discussion and conclusion}

We suggest that the observed diffuse GeV gamma-rays and the 511 keV annihilation flux from $5^{\circ}$ within the GC are consequences of multiple stellar captures by the Galactic black hole. The average injection energy carried away by relativistic protons is $\sim 6\times 10^{52}$ erg per capture every $10^5$ years. Such energy of the injection rate is sufficient to explain the observed electron/positron annihilation flux. 
  In our model, most positrons are produced in the central high density region (50 pc), and diffuse through the low density region to a distance $\sim 500$ pc during their thermalization time. Every %%@
capture takes place once every $10^5$ years. Therefore the observed positron annihilation emission results from a population of thermalized positrons, which are produced, cooled down and %%@
accumulated by past hundreds of capture events instead of a single injection process.
The gamma-ray intensity emitted from the central $0.5^{\circ}$ region is very sensitive %%@
in time and the current intensity
is most likely much weaker than its maximum value.
Our numerical calculations indicate that the in-flight annihilation cannot produce
enough 1-10 MeV gamma-rays to explain the observed data by COMPTEL. On the other hand,
unlike the diffuse gamma-rays in GeV range, the diffuse 1-10 MeV gamma-rays do not have a notable %%@
concentration within 500 pc. It is possible that they have different origins.

\begin{acknowledgements}
      The authors are grateful to A.~Aharonyan, E.~Churazov, Y.F. Huang, S. Komossa and F. %%@
Yuan  for very useful discussions, to Andy Strong who sent us
       unpublished COMPTEL data for the $5^{\circ} \times 5^{\circ}$ central region and
       to the anonymous referee for very useful comments. KSC is
supported by a RGC grant of Hong Kong Government, VAD and DOC were
partly supported by the NSC-RFBR Joint Research Project
95WFA0700088 and by the grant of a President of the Russian
Federation ``Scientific School of Academician V.L.Ginzburg".

\end{acknowledgements}


\begin{thebibliography}{58}
\expandafter\ifx\csname natexlab\endcsname\relax\def\natexlab#1{#1}\fi
\expandafter\ifx\csname url\endcsname\relax
  \def\url#1{{\tt #1}}\fi
\expandafter\ifx\csname urlprefix\endcsname\relax\def\urlprefix{URL }\fi

\bibitem[{Aharonian \& Neronov(2005{\natexlab{a}})}]{ahaner05a}
Aharonian F., Neronov A., 2005{\natexlab{a}}, Ap\&SS, 300, 255

\bibitem[{Aharonian \& Neronov(2005{\natexlab{b}})}]{ahaner05b}
Aharonian F., Neronov A., 2005{\natexlab{b}}, ApJ, 619, 306

\bibitem[{Aharonian et~al.(2004)Aharonian, Akhperjanian, Aye et~al.}]{ahaakh04}
Aharonian F., Akhperjanian A.G., Aye K.M., et~al., 2004, A\&A, 425, L13

\bibitem[{Bahcall \& Wolf(1976)}]{bahwol76}
Bahcall J.N., Wolf R.A., 1976, ApJ, 209, 214

\bibitem[{Barth et~al.(2004)Barth, Ho, Rutledge, \& Sargent}]{barho04}
Barth A.J., Ho L.C., Rutledge R.E., Sargent W.L.W., 2004, ApJ, 607, 90

\bibitem[{Beacom \& Y\"uksel(2006)}]{beayuk06}
Beacom J.F., Y\"uksel H., 2006, Phys. Rev. Lett., 97, 071102

\bibitem[{Beck(2007)}]{bec07}
Beck R., 2007, EAS Publications Series, 23, 19

\bibitem[{Berezinskii et~al.(1990)Berezinskii, Bulanov, Dogiel, \&
  Ptuskin}]{berbul90}
Berezinskii V.S., Bulanov S.V., Dogiel V.A., Ptuskin V.S., March 1990, In:
  Ginzburg V. (ed.) Astrophysics of cosmic rays, Amsterdam, North-Holland

\bibitem[{Cheng \& Lu(2001)}]{chelu01}
Cheng K.S., Lu Y., 2001, Mon. Not. R. Astron. Soc., 320, 235

\bibitem[{Cheng et~al.(2006)Cheng, Chernyshov, \& Dogiel}]{cheche06}
Cheng K.S., Chernyshov D.O., Dogiel V.A., 2006, ApJ, 645, 1138

\bibitem[{Churazov et~al.(2005)Churazov, Sunyaev, Sazonov, Revnivtsev, \&
  Varshalovich}]{chusun05}
Churazov E., Sunyaev R., Sazonov S., Revnivtsev M., Varshalovich D., 2005, Mon.
  Not. R. Astron. Soc., 357, 1377

\bibitem[{Cohn \& Kulsrud(1978)}]{cohkul78}
Cohn H., Kulsrud R.M., 1978, ApJ, 226, 1087

\bibitem[{Dogel \& Sharov(1985)}]{dogsha85}
Dogel V.A., Sharov G.S., 1985, Soviet Astronomy Letters, 11, 346

\bibitem[{Dogiel \& Ginzburg(1989)}]{doggin89}
Dogiel V.A., Ginzburg V.L., 1989, Space Sci. Rev., 49, 311

\bibitem[{Donley et~al.(2002)Donley, Brandt, Eracleous, \& Boller}]{donbra02}
Donley J.L., Brandt W.N., Eracleous M., Boller T., 2002, ApJ, 124, 1308

\bibitem[{Fender(2003)}]{Fen03}
Fender R., 2003, astro-ph/0303339

\bibitem[{Fender \& Maccarone(2004)}]{fenmac04}
Fender R., Maccarone T., 2004, In: Cheng K.S., Romero G.E. (eds.) Cosmic
  Gamma-Ray Sources, 205, Kluwer Academic Publishers, Dordrecht

\bibitem[{Fender et~al.(1999)Fender, Corbel, Tzioumis et~al.}]{fencor99}
Fender R., Corbel S., Tzioumis T., et~al., 1999, ApJ, 519, L165

\bibitem[{Ferrarese(2002)}]{fer02}
Ferrarese L., 2002, ApJ, 578, 90

\bibitem[{Gallo et~al.(2003)Gallo, Fender, \& Pooley}]{GalFen03}
Gallo E., Fender R.P., Pooley G.G., 2003, Mon. Not. R. Astron. Soc., 344, 60

\bibitem[{Gebhardt et~al.(2002)Gebhardt, Rich, \& Ho}]{gebric02}
Gebhardt K., Rich R.M., Ho L.C., 2002, ApJ, 578, L41

\bibitem[{Genzel \& Karas(2007)}]{genkar07}
Genzel R., Karas V., 2007, arXiv:0704.1281

\bibitem[{Halpern et~al.(2004)Halpern, Gezari, \& Komossa}]{halgez04}
Halpern J.P., Gezari S., Komossa S., 2004, ApJ, 604, 572

\bibitem[{Han et~al.(2006)Han, Manchester, Lyne, Qiao, \& van
  Straten}]{hanman06}
Han J.L., Manchester R.N., Lyne A.G., Qiao G.J., van Straten W., 2006, ApJ,
  642, 868

\bibitem[{Hills(1975)}]{hil75}
Hills J.G., 1975, Nat, 254, 295

\bibitem[{Jean et~al.(2006)Jean, Kn\"odlseder, Gillard et~al.}]{jeakno06}
Jean P., Kn\"odlseder J., Gillard W., et~al., 2006, A\&A, 445, 579

\bibitem[{Komossa(2006)}]{kom06}
Komossa S., 2006, private communication

\bibitem[{Kosack et~al.(2004)Kosack, Badran, Bond et~al.}]{kosbad04}
Kosack K., Badran H.M., Bond I.H., et~al., 2004, ApJ, 608, L97

\bibitem[{Lightman \& Shapiro(1977)}]{ligsha77}
Lightman A.P., Shapiro S.L., 1977, ApJ, 211, 244

\bibitem[{Lu et~al.(2006)Lu, Cheng, \& Huang}]{luche06}
Lu Y., Cheng K.S., Huang Y.F., 2006, ApJ, 641, 288

\bibitem[{Magorrian \& Tremaine(1999)}]{magtre99}
Magorrian J., Tremaine S., 1999, Mon. Not. R. Astron. Soc., 309, 447

\bibitem[{Markoff et~al.(1997)Markoff, Melia, \& Sarcevic}]{marmel97}
Markoff S., Melia F., Sarcevic I., 1997, ApJ, 489, L47

\bibitem[{Mastichiadis \& Ozernoy(1994)}]{masoze94}
Mastichiadis A., Ozernoy L.M., 1994, ApJ, 426, 599

\bibitem[{Mayer-Hasselwander et~al.(1998)Mayer-Hasselwander, Bertsch, Dingus
  et~al.}]{mayber98}
Mayer-Hasselwander H.A., Bertsch D.L., Dingus B.L., et~al., 1998, A\&A, 335,
  161

\bibitem[{Melia \& Falcke(2001)}]{melfal01}
Melia F., Falcke H., 2001, ARA\&A, 39, 309

\bibitem[{Morfill(1982)}]{mor82}
Morfill G.E., 1982, ApJ, 262, 749

\bibitem[{Oka \& Manmoto(2003)}]{okaman03}
Oka K., Manmoto T., 2003, Mon. Not. R. Astron. Soc., 340, 543

\bibitem[{Padoan \& Scalo(2005)}]{padsca05}
Padoan P., Scalo J., 2005, ApJ, 624, L97

\bibitem[{Parizot et~al.(2005)Parizot, Cass\'e, Lehoucq, \& Paul}]{parcas05}
Parizot E., Cass\'e M., Lehoucq R., Paul J., 2005, A\&A, 432, 889

\bibitem[{Phinney(1989)}]{phi89}
Phinney E.S., 1989, Nat, 340, 595

\bibitem[{Pohl(1997)}]{poh97}
Pohl M., 1997, A\&A, 317, 441

\bibitem[{Prantzos(2006)}]{pra06}
Prantzos N., 2006, A\&A, 449, 869

\bibitem[{Rees(1988)}]{ree88}
Rees M.J., 1988, Nat, 333, 523

\bibitem[{Skilling \& Strong(1976)}]{skistr76}
Skilling J., Strong A.W., 1976, A\&A, 53, 253

\bibitem[{Strong \& Moskalenko(1998)}]{strmos98}
Strong A.W., Moskalenko I.V., 1998, ApJ, 509, 212

\bibitem[{Strong et~al.(2000)Strong, Moskalenko, \& Reimer}]{strmos00}
Strong A.W., Moskalenko I.V., Reimer O., 2000, ApJ, 537, 736

\bibitem[{Strong et~al.(2005)Strong, Diehl, Halloin et~al.}]{strdie05}
Strong A.W., Diehl R., Halloin H., et~al., 2005, A\&A, 444, 495

\bibitem[{Syer \& Ulmer(1999)}]{syeulm99}
Syer D., Ulmer A., 1999, Mon. Not. R. Astron. Soc., 306, 35

\bibitem[{Totani(2006)}]{tot06}
Totani T., 2006, PASJ, 58, 965

\bibitem[{Tremaine et~al.(2002)Tremaine, Gebhardt, Bender et~al.}]{tregeb02}
Tremaine S., Gebhardt K., Bender R., et~al., 2002, ApJ, 574, 740

\bibitem[{Tsuchiya et~al.(2004)Tsuchiya, Enomoto, Ksenofontov
  et~al.}]{tsueno04}
Tsuchiya K., Enomoto R., Ksenofontov L.T., et~al., 2004, ApJ, 606, L115

\bibitem[{Wang et~al.(2006)Wang, Pun, \& Cheng}]{wanpun06}
Wang W., Pun C.S.J., Cheng K.S., 2006, A\&A, 446, 943

\bibitem[{Weekes(2004)}]{wee04}
Weekes T.C., 2004, In: Cheng K.S., Romero G.E. (eds.) Cosmic Gamma-Ray Sources,
  345, Kluwer Academic Publishers, Dordrecht

\bibitem[{Weidenspointner et~al.(2006)Weidenspointner, Shrader, Kn\"odlseder
  et~al.}]{weishr06}
Weidenspointner G., Shrader C.R., Kn\"odlseder J., et~al., 2006, A\&A, 450,
  1013

\bibitem[{Wong et~al.(2007)Wong, Huang, \& Cheng}]{wonhua07}
Wong A.Y.L., Huang Y.F., Cheng K.S., 2007, accepted by A\&A (arXiv:0704.3480)

\bibitem[{Yuan(2007)}]{yua07}
Yuan F., 2007, private communication

\bibitem[{Yuan et~al.(2002)Yuan, Markoff, \& Falcke}]{yuamar02}
Yuan F., Markoff S., Falcke H., 2002, A\&A, 383, 854

\bibitem[{Yuan et~al.(2005)Yuan, Cui, \& Narayan}]{yuacui05}
Yuan F., Cui W., Narayan R., 2005, ApJ, 620, 905

\end{thebibliography}
\end{document}